\documentclass[prd,preprint,aps,unsortedaddress]{revtex4}

\usepackage{amsfonts}
\usepackage{amssymb}
\usepackage{amsmath}
\usepackage[all]{xy}
\newcommand{\be}{\begin{equation}}
\newcommand{\ee}{\end{equation}}
\newcommand{\bn}{\begin{eqnarray}}
\newcommand{\en}{\end{eqnarray}}

\begin{document}

\title{Radiative processes as a condensation phenomenon and the physical meaning of deformed canonical structures}
\author{J. Gamboa}
\email{jgamboa@usach.cl}
\affiliation{Departamento de F\'{\i}sica, Universidad de Santiago de Chile,
Casilla 307, Santiago, Chile}
\author{L. S. Grigorio}
\email{leogrigorio@if.ufrj.br}
\affiliation{Instituto de F\'\i sica, Universidade
Federal do Rio de Janeiro, 21945, Rio de Janeiro, Brazil}
\author{M. S. Guimaraes}
\email{marcelosg@if.ufrj.br} \affiliation{Instituto de F\'\i sica, Universidade
Federal do Rio de Janeiro, 21945, Rio de Janeiro, Brazil}
\author{F. M\'endez}
\email{fmendez@usach.cl}
\affiliation{Departamento de F\'{\i}sica, Universidad de Santiago de Chile,
Casilla 307, Santiago, Chile}
\author{C. Wotzasek}
\email{clovis@if.ufrj.br}
\affiliation{Instituto de F\'\i sica, Universidade Federal do Rio de
Janeiro, 21945, Rio de Janeiro, Brazil}
\date{\today}

\begin{abstract}

Working with well known models in $(2+1)D$ we discuss the physics behind the deformation of the canonical structure of
these theories. A new deformation is constructed linking the massless scalar field theory with the self-dual theory. This is the
exact dual of the known deformation connecting the Maxwell theory with the Maxwell-Chern-Simons theory. Duality is used to
establish a web of relations between the mentioned theories and a physical picture of the deformation procedure is
suggested.

\end{abstract}

\maketitle


\section{Introduction}

In this work we will construct a web of relations connecting known theories in
$(2+1)D$ through duality and deformation. By deformation we mean a modification of
the canonical structure of a theory. The deformation procedure, unlike duality,
connects theories with different physical content. Therefore it is reasonable to
suggest that the seemingly \emph{ad hoc} deformations are in fact emulating some
physical process. It is the ultimate purpose of this work to give a more precise
meaning to this statement. Further, through the duality connection, we will be able
to draw important conclusions concerning the duality of the physical processes
themselves.

The deformation can be interpreted as defining a quantum theory with noncommutative fields as was
introduced by \cite{Carmona:2002iv,Carmona:2003kh} and further studied by
\cite{Mandanici:2004ht,Balachandran:2007ua,Das:2005uq,Gamboa:2005pd}. This is a different set from the maybe more familiar
spacetime noncommutativity, where the coordinates of spacetime are required to satisfy
algebraic relations such as
\begin{eqnarray}
\label{01}
 [x_\mu, x_\nu] = i \theta_{\mu\nu}
\end{eqnarray}
with $\theta^{\mu\nu}$ a $(lenght)^2$ parameter. In this way equation (\ref{01})
defines a modification of the spacetime structure suppressing absolute spacetime
localization by setting up a minimal area scale. In the noncommutative fields
framework on the other hand, the modification is on the structure of the field phase
space, for example, in $(3+1)D$ one would have
\begin{eqnarray}
\label{02}
\left[\Phi_i(x), \Phi_j(y)\right] = i \varepsilon_{ijk} B^k \delta^{(3)}(x-y)\nonumber\\
\left[\Phi_i(x), \Pi_j(y)\right] = i \delta_{ij} \delta^{(3)}(x-y) \nonumber\\
\left[\Pi_i(x), \Pi_j(y)\right] = i \varepsilon_{ijk} \Theta^k \delta^{(3)}(x-y).
\end{eqnarray}
Where $\Pi$ is the canonical momentum. The parameters $B$ and $\Theta$ appearing in
(\ref{02}) introduce new scales to the problem. They have canonical dimensions of
length and mass, respectively, introducing an ultra-violet and an infra-red scale
respectively if they are taken to be small. In $(3+1)D$ these deformations
potentially break Lorentz and CPT invariance. As will be seen, in $(2+1)D$ example treated
here they break $P$ and $T$.

The algebraic relations (\ref{02}) are viewed as deformations of the canonical
structure defining the relevant field theory. As such they lead to different
theories describing distinct and new physical phenomena parameterized by the
continuous deformation parameters.

The widespread applications of these ideas rely on the fact that these deformations
can be taken to lead to only tiny observed effects in known physical theories.
Since there seems to be no inconsistences in their formulations they constitute a
promising fertile ground for the study of new physics beyond presently known models.
Therefore the rationale of the procedure deserves further inquire. The results
discussed here can be seen also as a consistency check of the whole deformation
procedure since it will be shown that it behaves appropriately under well known
duality relations.

A great deal of work  has been dedicated to the investigation of the physical,
observable consequences of these controlled modifications in $(3+1)D$. The violation
of Lorentz invariance following from these deformed structures leads to many
interesting new phenomena. We may mention a few: in
\cite{Carmona:2002iv,Carmona:2003kh} it was shown that, considering this deformation
as modeling the high energy cosmic-ray radiation, it is possible to choose the
deformation parameter such as to turn on or off the interaction of the cosmic-ray
with the cosmic background radiation thus circumventing a GZK cutoff. In
\cite{Arias:2007zz} it was shown that neutrino oscillations may be achieved even for
massless neutrinos if we allow for Lorentz violation as defined by these
deformations. In \cite{Carmona:2004xc} baryogenesis was suggested as a possibility
even in termal equilibrium, bypassing one of the Sakharov's criteria, if CPT is
broken as it is by the deformation. Recently \cite{Barosi:2008gx} inflationary
scenarios were investigated in this context too, the deformation of the canonical
structure is this case being defined not locally by a delta function as in
(\ref{02}) but in a region parameterized by a further adjustable parameter and the
different scenarios were studied as a function of this parameter. This modified
deformation was also discussed in \cite{Balachandran:2007ua}.

These phenomenological consequences follow from the deformed induced models. It is
this general interpretation of the deformation as a model generator that is of
interest to us in the present work. In what follows we will stick with a particular
example where all models related by duality and deformation are well known and
studied. This will allow us to inquire not only about the consistency of the
deformation procedure but also to obtain important new results concerning the
relations between these models.

We will specifically address a $(2+1)D$ situation, a framework in which the
deformation can be done without breaking the Lorentz symmetry but breaking $P$ and
$T$ symmetries. We will demonstrate that the Self-dual (SD) model \cite{tpn} can be obtained by a deformation of
the canonical structure of the free
scalar theory. This is a new result that displays the very interesting property of
``rank jump'', that is, the scalar theory is turned into a vector field theory after
deformation. Incidentally this is the exact dual manifestation of the result
discussed in \cite{petrov}. There it was shown that a deformation of the canonical
structure of the free Maxwell theory gives the Maxwell-Chern-Simons (MCS) theory
\cite{Deser:1981wh} as a result (no ``rank jump'' here). Duality shows up in the fact that in
$(2+1)D$ the free Maxwell theory has a dual description in terms of the free scalar
theory and the MCS theory is well known to be dual to the SD theory \cite{dj}.

With the result presented here we have then a neat picture connecting those four
models through deformation and duality \be 
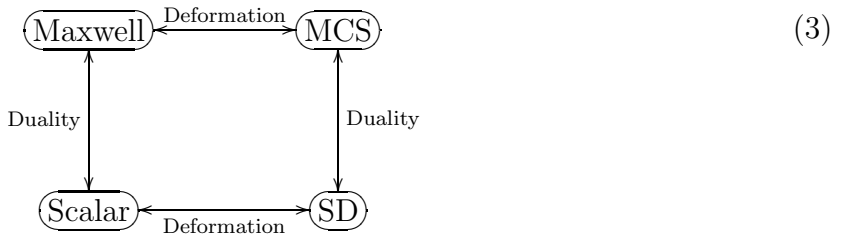
\label{diagrama}
 \xymatrix{*+[F-:<10pt>]{\rm{Maxwell}}+\ar[rr]^-{\rm{Deformation}}\ar[dd] & &
*+[F-:<10pt>]{\rm{MCS}} \ar[ll]\ar[dd]\\
 &  & \\
*+[F-:<10pt>]{\rm{Scalar}} \ar[uu]^-{\rm{Duality}} \ar[rr]_-{\rm{Deformation}}& &
*+[F-:<10pt>]
{\rm{SD}}  \ar[uu]_-{\rm{Duality}}\ar[ll]\\
         }
\ee

As already pointed out, another very important point which certainly deserves
special attention is the possible physical interpretation of the deformation
procedure. What it means, physically, an ordinary field theory to turn into a
noncommutative fields theory? The upper link of the diagram (\ref{diagrama}),
connecting the Maxwell theory with the MCS theory, has a well known physical
interpretation: it describes the introduction of massive fermions in the theory
which gives rise to vacuum polarization effects inducing the Chern-Simons term
through radiative corrections. We think that one of the most important results of
our work is to give a physical picture of what is going on in the lower link of the
diagram, the scalar-SD connection. Because of the duality links, it can be
interpreted as the dual phenomenon to the fermionic radiative corrections.

In the next section we will review the known connections in this picture: the
Maxwell-scalar duality and the MCS-SD duality, as well as the deformation of the
Maxwell theory leading to the MCS theory. In the next following section we will
present our result demonstrating the scalar-SD connection thus completing the
picture (\ref{diagrama}) and establishing it as the dual manifestation of the
Maxwell-MCS connection. We then proceed to a discussion of the physics underlying
this picture which is the other result of this work. After that we finish with our
concluding remarks.

\section{Duality and deformation}

The purpose of this section is to review some known results that are
relevant to our work.

\subsection{Duality}

The two duality connections figuring in (\ref{diagrama}) share the same basic
principles, it can be viewed in some sense as a change of variables. The duality
procedure consists in doing a sequence of manipulations in the action making sure
that none of them alter the physical content of the theory. Let us start with the
Maxwell-Chern-Simons theory
\begin{eqnarray}
  S_{MCS}= \int d^3x \left\{-\frac 14 F_{\mu\nu}F^{\mu\nu} - \frac m2 A_{\mu}\varepsilon^{\mu\nu\rho}\partial_{\nu}A_{\rho}
  \right\},
  \label{mcsad01}
\end{eqnarray}
where $F_{\mu\nu} = \partial_{\mu}A_{\nu} - \partial_{\nu}A_{\mu}$. It describes a
topologically massive degree of freedom and formally reduces to the pure Maxwell
theory if the topological mass $m \rightarrow 0$. Under an abelian gauge
transformation $S_{MCS}$ changes by a surface term, so if it is defined in a suitable
space-time (topologically trivial and boundaryless) it is gauge invariant. But there
is a remarkable property concerning this gauge symmetry structure: it is completely
independent of the dynamical content of the theory. Thus when we talk about duality
we are referring to dynamical duality. This is easily seen as follows (the following
discussion is taken from \cite{Guimaraes:2006gj}
where a somewhat more complete assessment is made)

Consider the Lagrangian obtained from the MCS lagrangian by the introduction of an
auxiliary field
\begin{eqnarray}
 {\cal L}_{M} &=& \Pi_{\mu}(\varepsilon^{\mu\nu\rho}\partial_{\nu}A_{\rho}) + \frac 12 \Pi_{\mu}\Pi^{\mu} - \frac m2 A_{\mu}
 \varepsilon^{\mu\nu\rho}\partial_{\nu}A_{\rho}\nonumber\\
              &=&(\Pi_{\mu} - \frac m2 A_{\mu})(\varepsilon^{\mu\nu\rho}\partial_{\nu}A_{\rho}) + \frac 12 \Pi_{\mu}\Pi^{\mu},
    \label{mcsadmestra}
\end{eqnarray}
where $\Pi_\mu$ is an auxiliary vector-field which may be integrated out to give us
back the MCS lagrangian. By making the redefinition $\Pi_{\mu} - \frac m2 A_{\mu} =
B_{\mu}$, we find
\begin{eqnarray}
{\cal L}_{M} = B_{\mu}(\varepsilon^{\mu\nu\rho}\partial_{\nu}A_{\rho}) + \frac 12
(B_{\mu} + \frac m2 A_{\mu})(B^{\mu} + \frac m2 A^{\mu}).
    \label{mcsad011}
\end{eqnarray}
Observe that by definition $B_{\mu}$ transforms as $B_{\mu} \rightarrow B_{\mu} -
\frac m2 \partial_{\mu}\Lambda$ whenever $A_{\mu} \rightarrow A_{\mu} +
\partial_{\mu}\Lambda$, so that the gauge character of the $A_\mu$ field did not changed.
Next we may perform a canonical transformation in the space of the fields to reveal
the self-dual and pure gauge nature of the components
\begin{eqnarray}
 B_{\mu} &=& \frac 12 (A^{+}_{\mu} - A^{-}_{\mu})\nonumber\\
   A_{\mu} &=& \frac 1m (A^{+}_{\mu} + A^{-}_{\mu}),
    \label{mcsad022}
\end{eqnarray}
which gives us
\begin{eqnarray}
{\cal L}_{M} = \frac{1}{2m}
A^{+}_{\mu}\varepsilon^{\mu\nu\rho}\partial_{\nu}A^{+}_{\rho} - \frac{1}{2m}
A^{-}_{\mu}\varepsilon^{\mu\nu\rho}\partial_{\nu}A^{-}_{\rho} + \frac 12
A^{+}_{\mu}A^{+\mu},
    \label{mcsad03}
\end{eqnarray}
or, renaming $A^{+}_{\mu} = f_{\mu}$ and $A^{-}_{\mu} = A_{\mu}$
\begin{eqnarray}
 {\cal L}_{M} = -\frac{1}{2m} A_{\mu}\varepsilon^{\mu\nu\rho}\partial_{\nu}A_{\rho} +  \frac 12 f_{\mu}f^{\mu} + \frac{1}{2m} f_{
 \mu}\varepsilon^{\mu\nu\rho}\partial_{\nu}f_{\rho}  = {\cal L}_{CS} + {\cal L}_{SD},
    \label{mcsad04}
\end{eqnarray}
The first term is the standard Chern-Simons lagrangean and the
remaining terms we recognize as the SD model \cite{tpn}. It is clear by this
procedure that the kind of gauge symmetry carried by the MCS theory have completely
innocuous dynamical character as is well know by the properties of the pure
Chern-Simons theory \cite{Dunne:1998qy}. As we were able to separate this term we might
say that the only information that the MCS theory has, which is not present in the
SD model, regards the topological character of the space in which the theory is
defined. It is very interesting and highly nontrivial that this separation is
possible. It means, for example, that the energy propagating modes is given entirely
by the SD sector in (\ref{mcsad04}) but the energy eigenstates have a degeneracy
parameterized by the Hilbert space of pure Chern-Simons theory (a topological
degeneracy) \cite{Asorey:1993ft}. In this way we establish the MCS-SD dynamical duality
connection depicted in the diagram (\ref{diagrama}).

Incidentally eq.(\ref{mcsad01}) describes the Maxwell theory if $m=0$. In this case,
performing a partial integration, it is clear from (\ref{mcsadmestra}) that
$A_{\mu}$ appears as a Lagrange multiplier which enforces the constraint
\begin{eqnarray}
 \varepsilon^{\mu\nu\rho}\partial_{\nu}\Pi_{\rho} = 0 \Rightarrow \Pi_{\mu} = \partial_{\mu}\phi
    \label{contr}
\end{eqnarray}
where $\phi$ is a scalar field. Substituting back in the action we obtain a
free scalar theory, which proves the scalar-Maxwell duality connection as well.

\subsection{Deformation}

In this section we will reobtain the results of \cite{petrov} regarding the
connection between the free Maxwell theory and the MCS theory through the
deformation of its canonical structure.
The usual analysis gives rise to the following structure,
consisting of the Hamiltonian,
\begin{eqnarray}
H = \int d^2x \left[ \frac 12 \vec{\Pi}^2 + \frac 12 \left(\nabla \times
\vec{A}\right)^2\right],
    \label{hamilmax}
\end{eqnarray}
the canonical brackets
\begin{eqnarray}
\label{canmax}
\left\{A_i(x), A_j(y)\right\} &=& 0\nonumber\\
\left\{A_i(x), \Pi^j(y)\right\} &=&  \delta^j_i \delta^{(2)}(x-y) \nonumber\\
\left\{\Pi^i(x), \Pi^j(y)\right\} &=& 0 \, ,
\end{eqnarray}
and the constraint
\begin{eqnarray}
\label{constmax} \nabla\cdot\vec{\Pi}=0
\end{eqnarray}
which could be incorporated in the definition of the Hamiltonian through a Lagrange
multiplier which turns out to be the scalar potential $A^0$. The constraint reduces
the phase space to a two dimensional manifold leaving the
system with one degree of freedom as is known to be the case for the Maxwell theory
in $(2+1)D$. The equations of motion, along with the constraint and the
identifications $\vec{E}= \vec{\Pi}$ and $B = \nabla \times \vec{A}$ for the
electric and magnetic fields respectively, gives us the Maxwell's equation in
$(2+1)D$ (note that $B$ is a pseudo-scalar).

Now we turn to the deformation procedure. A deformation of the canonical
brackets is an explicit modification of the dynamics of the theory. It is well known
that in $(2+1)D$ a Chern-Simons (CS) term can be dynamically induced in the
effective gauge theory by fermionic interactions. This can be inferred from symmetry
considerations: if fermions are massive the effective gauge theory must break the
discrete $P$ and $T$ symmetries since a fermionic mass does it in $(2+1)D$. But even if
the fermions are massless the effective gauge theory breaks $P$ and $T$
because it is impossible to maintain gauge symmetry along with the $P$ and $T$
symmetries after regularization. So $P$ and $T$ are broken once we recognize that gauge
symmetry, being just a redundancy in our description of the system, cannot disappear
due to dynamical effects. We will take this recognition of the fundamental role of
gauge symmetry as a guiding principle in obtaining the resulting theory after
deformation.

As was done in \cite{petrov} (see also \cite{Gamboa:2005bf}) we will impose the following
deformation to the brackets (\ref{canmax})
\begin{eqnarray}
\label{defcanmax}
\left\{A_i(x), A_j(y)\right\} &=& 0\nonumber\\
\left\{A_i(x), \Pi^j(y)\right\} &=&  \delta^j_i \delta^{(2)}(x-y) \nonumber\\
\left\{\Pi^i(x), \Pi^j(y)\right\} &=& \varepsilon^{ij}m\delta^{(2)}(x-y),
\end{eqnarray}
where $m$, if taken small, is interpreted as an IR deformation (as it then
defines a low energy scale for the deformation).

Alone, this deformation breaks the gauge invariance as the hamiltonian no longer has
vanishing brackets with the original gauge generator
\begin{eqnarray}
\label{gaugegen} G[\alpha] = \int d^2x \alpha(x) \nabla\cdot\vec{\Pi}.
\end{eqnarray}
This prompt us to modify this generator defining a new constraint \cite{Gamboa:2005bf}
\begin{eqnarray}
\label{gaugegendef} \tilde{G}[\alpha] = \int d^2x \alpha(x) \partial_i\left(\Pi^i -
\varepsilon^{ij}m A_j\right).
\end{eqnarray}
which may be seen to have vanishing brackets with the Hamiltonian (\ref{hamilmax}). This new
generator is obtained in a unique way just by requiring it to have vanishing
brackets with the canonical momentum. To further construct the Lagrangian it is
better to put the deformed brackets in a canonical form from which the symplectic
structure can be immediately recognized. This is done by the following redefinition
of the momenta
\begin{eqnarray}
\label{redefmomenta} \tilde{\Pi}^i = \Pi^i - \frac{\varepsilon^{ij}m}{2} A_j
\end{eqnarray}
which cast the deformed brackets in the form of canonical ones
\begin{eqnarray}
\label{defcanmax}
\left\{A_i(x), A_j(y)\right\} &=& 0\nonumber\\
\left\{A_i(x), \tilde{\Pi}^j(y)\right\} &=&  \delta^j_i \delta^{(2)}(x-y) \nonumber\\
\left\{\tilde{\Pi}^i(x), \tilde{\Pi}^j(y)\right\} &=& 0 \, .
\end{eqnarray}
The Lagrangian is then easily constructed if we take care of the new constraint also
(here ${\cal H}$ is the Hamiltonian density corresponding to (\ref{hamilmax}))
\begin{eqnarray}
\label{deflag} {\cal L} &=& \sum_i \tilde{\Pi}^i\dot{A}_i - {\cal H} + A^0
\partial_i\left(\tilde{\Pi}^i - \varepsilon^{ij}m
A_j\right)\nonumber\\
&=& \sum_i \left(\Pi^i\dot{A}_i  +
\frac{m}{2}\varepsilon^{ij}A^i\dot{A}_j\right) - \frac 12 \vec{\Pi}^2 - \frac
12 \left(\nabla \times \vec{A}\right)^2 + A^0
\partial_i\left(\Pi^i -
\varepsilon^{ij}m A_j\right)\nonumber\\
&=& - \frac14 F^{\mu\nu} F_{\mu\nu} - \frac{m}{2}
\varepsilon^{\mu\nu\rho}A_{\mu}\partial_{\nu}A_{\rho}
\end{eqnarray}
where the last equality is obtained from using the identification $\Pi^i = \dot{A}_i
- \partial_i A^0$. This establishes the connection Maxwell/MCS through deformation as
depicted in the diagram (\ref{diagrama}). Next we will complete the picture.

\section{The free scalar / self-dual connection}

The picture depicted in (\ref{diagrama}) that have been drawn in the previous
sections will be completed now. In this section we will establish the scalar/SD
connection through deformation. This is to be viewed as the exact dual of the
procedure discussed in the last section. But here the interesting property of ``rank jump''
will show up. The deformation considered here will turn a scalar field into
a vector field. This is in tune with the scalar/Maxwell duality of course, but
resulting from the deformation rather than of duality it may have a deeper meaning
originating from new dynamical effects. These new effects would therefore be the
exact dual of the radiative corrections emulated by the Maxwell/MCS connection
discussed in the last section. We will have more to say about this in the concluding
remarks.

Consider the free scalar theory defined by the Hamiltonian
\begin{eqnarray}
H = \int d^2x \left[ \frac 12 \Pi^2 + \frac 12
\left(\nabla\cdot\phi\right)^2\right],
    \label{hamilsca}
\end{eqnarray}
and the canonical brackets
\begin{eqnarray}
\label{cansca}
\left\{\phi(x), \phi(y)\right\} &=& 0\nonumber\\
\left\{\phi(x), \Pi(y)\right\} &=& \delta^{(2)}(x-y) \nonumber\\
\left\{\Pi(x), \Pi(y)\right\} &=& 0.
\end{eqnarray}
We propose the following deformation to the brackets
\begin{eqnarray}
\label{defcansca}
\left\{\partial_i\phi(x), \partial_j\phi(y)\right\} &=& m\varepsilon_{ij}\delta^{(2)}(x-y)\nonumber\\
\left\{\phi(x), \Pi(y)\right\} &=& \delta^{(2)}(x-y) \nonumber\\
\left\{\Pi(x), \Pi(y)\right\} &=& 0.
\end{eqnarray}
and claim that the resulting system defined by (\ref{hamilsca}) and
(\ref{defcansca}) is equivalent to the SD system defined by the action:
\begin{eqnarray}
 {\cal L}_{SD} = \frac 12 f_{\mu}f^{\mu} + \frac{1}{2m} f_{\mu}\varepsilon^{\mu\nu\rho}\partial_{\nu}f_{\rho}\, .
    \label{sd}
\end{eqnarray}
In order to establish this link consider the equations of motion of the deformed
scalar system
\begin{eqnarray}
\label{scadefeqmot}
\dot{\Pi} &=& \nabla^2\phi  \nonumber\\
\partial_i\dot{\phi} &=& \partial_i\Pi - m\varepsilon_{ij}\partial^j\phi.
\end{eqnarray}
These equations contain only derivatives of the scalar field. It must be further
noticed that, if $m \neq 0$, these equations will have only solutions of the
form $\phi=f(t)$ and $\Pi = c$, with $f$ a function of time only and $c$ a constant,
both to be determined by initial conditions. This is then a trivial system with an homogeneous
space energy density. But if we assign no meaning to the scalar field
itself, claiming that its derivatives must be taken as the fundamental dynamical
fields we are able to construct a non-trivial system. We will later elaborate a bit
more on the possible physical phenomena behind this procedure, for now we will
simply define the new dynamical variables for this system by the map
\begin{eqnarray}
\label{scasdmap}
\Pi &=& f^0 \nonumber\\
\partial^i \phi &=& f^i.
\end{eqnarray}
Equations (\ref{scadefeqmot}), in terms of these new variables, can be cast in the
form
\begin{eqnarray}
\label{scasdeq}
\partial_{\mu}f^{\mu} &=& 0  \nonumber\\
f^i + \frac{1}{m}\varepsilon^{ij}\partial_j f_0 -
\frac{1}{m}\varepsilon^{ij}\partial_0 f_j &=& 0.
\end{eqnarray}
Observe further that by taking the derivative of the second equation in
(\ref{scasdeq}) and using the first we have
\begin{eqnarray}
\label{sdconst} \partial_i f^i =
\frac{1}{m}\partial_0\left(\varepsilon^{ij}\partial_i f_j\right) =
-\partial_0f^0
\end{eqnarray}
Which allows us to conclude that
\begin{eqnarray}
\label{sdconst2}  f^0 = -\frac{1}{m}\varepsilon^{ij}\partial_i f_j + C(x,y)
\end{eqnarray}
That is, the field $f^0$  has its dynamics determined by the fields $f_i$ except for
a time independent function, $C(x,y)$, which may be absorbed in a redefinition of
$f^0$ without modifying the dynamics. By doing that we may compactly write the
information contained in the above equations (\ref{scasdeq}) and (\ref{sdconst2}) in
the form
\begin{eqnarray}
\label{sdeq} f^{\mu} + \frac{1}{m}\varepsilon^{\mu\nu\rho}\partial_{\nu}
f_{\rho} = 0
\end{eqnarray}
which is recognized as the Euler-Lagrange equations for the $SD$ field whose action
is defined by (\ref{sd}). This proves therefore that all non-trivial dynamical content
of the deformed scalar field system is reproduced by the $SD$ system.

\section{The physical picture}

In this section we will discuss the physical meaning of the connections just
established in the previous sections.
The diagram (\ref{diagrama}) we have constructed contain four links. The two duality
relations are well known, as discussed in section II.A, so are their physical
interpretation. They are a statement of the fact that the system admits an
equivalent description in terms of another set of variables.

This equivalent description is much sought after in fully interacting theories
where, by the general properties of duality, it would allow for a perturbative,
weakly coupled, description of strongly interacting theories. The general
implementation of duality in fully interacting theories is a remarkably difficult
task though, with only very special examples known \cite{AlvarezGaume:1997ix}. In our diagram, the
duality links connects free theories but it can be readily generalized to include
couplings with external sources. As we will see further ahead in this section, this
is already sufficient to construct a picture of the physical mechanism behind the
deformation.

The upper link connects the Maxwell theory with the MCS theory. It is known
\cite{Redlich:1983dv} that if massive fermions are taken to interact with the Maxwell field in
$(2+1)D$ the effective description, after ``integrating out'' the fermions, is the
MCS theory to lowest order in the inverse fermion mass. This is an excellent
approximation for the system if the fermions are heavy. The induced CS term has a
nice physical interpretation. If the fermions are much heavier than the other
relevant particles in the system they cannot be excited as real particles, their
dynamics are frozen, but they will contribute quantum mechanically through fermionic
loops which will surely disturb the propagation of the lighter particles. But
massive fermions have very peculiar properties in $(2+1)D$ which is intimately
related to the nature of spin in this dimensionality. In $(2+1)D$ a Dirac fermion
has only two components as follows from the representation of the Dirac matrices in
terms of the $2\times 2$ Pauli matrices. Therefore electrons have a definite spin
component, they will ``point up or down'', and this is defined by the sign of the
mass term in the action, which properly violates P and T symmetries. As a
consequence it follows that the fermionic quantum condensate in which the lighter
particles propagate is a $P$ and $T$ symmetry breaking state and the effective
description of the electromagnetic propagation must include a term with this
information, that is, the CS term.

In fact, the CS term may be viewed as an effective interaction (representing the
self energy of the fermion; its mass term) of the electromagnetic current coupled
with an induced current. Explicitly:
\begin{eqnarray}
\label{induced1} < M \bar{\psi}\psi> = <J^{\mu}> A_{\mu}
\end{eqnarray}
where the induced current takes the form
\begin{eqnarray}
\label{induced2}   <J^{\mu}> = \frac{e^2}{4\pi}\frac{M}{|M|} \ast F^{\mu}
\end{eqnarray}
Except for the $4\pi$ factor, this expression can be obtained only on the grounds
of dimensional analysis.
The preceding discussion of the fermionic condensate is essentially a reproduction
of remarks found in \cite{Redlich:1983dv} and \cite{Wilczek:1990ik}. It contains the essential
ingredients for an interpretation of the dual formulation.

The question is: how this physical picture presents itself in the dual formulation?
In the original picture just discussed we may think of the deformation as describing
a condensation process of sources minimally coupled with the electromagnetic field.
Consider the Maxwell field minimally coupled with an external conserved current
\begin{eqnarray}
\label{Mminimally}   {\cal L}_{Max} = -\frac 14 F^{\mu\nu}F_{\mu\nu}
-eJ_{\mu}A^{\mu},
\end{eqnarray}
and the conserved current can be written as
\begin{eqnarray}
\label{conscurre}  J^{\mu} = \varepsilon^{\mu\nu\rho}\partial_{\nu}K_{\rho} \, .
\end{eqnarray}
In doing this we are introducing a symmetry in the theory which accounts for the
freedom that we have in choosing the brane $K_{\mu}$
\begin{eqnarray}
\label{branesymm}  K_{\mu} \rightarrow K_{\mu} + \partial_{\mu} \chi
\end{eqnarray}
In this formulation $J^{\mu}$ describes a delta-like current which borders the
surface traced by the brane $K_{\mu}$. The condensation
process described previously thus can be seen as the phenomenon of the singular
brane $K_{\mu}$ turning into a field which in a first approximation (in a momentum
expansion) is identified with the gauge field $A_{\mu}$.

A dual representation of (\ref{Mminimally}) is constructed following the same path
traced in section II.A. The result is the scalar field non-minimally coupled with
the brane $K_{\mu}$
\begin{eqnarray}
\label{SNM}   {\cal L}_{Sc} = \frac 12 \left( \partial_{\mu} \phi - eK_{\mu}
\right)^2.
\end{eqnarray}
The brane symmetry is realized here taking into account the compactness of the
scalar field
\begin{eqnarray}
\label{branesymm2}  K_{\mu} &\rightarrow& K_{\mu} + \partial_{\mu} \chi,\nonumber\\
                    \phi &\rightarrow& \phi - e\chi.
\end{eqnarray}
Thus we see that the minimally coupled sources of the original formulation presents
themselves as defects in the dual formulation. Therefore the physical picture in the
dual formulation is a defect condensation.

To proceed further we refer to a very useful procedure to deal with this phenomenon
in this formulation, it is called the Julia-Toulouse mechanism. It was
further investigated and extended to deal with general relativistic p-form theories
in the presence of defects in a seminal work by Quevedo and Trugenberger
\cite{Quevedo:1996uu}. The prescription they have proposed was able to deliver a very interesting
picture of the dual phenomenon to the Higgs mechanism and they were also able to
draw important results concerning the puzzle of the axion mass.

The Julia-Toulouse prescription (or mechanism) amounts to the construction of a
field theory in which the defects are condensed. The details about the phenomenon
that drives the condensation are not addressed nor they need to be because with very
general assumptions an unique form of the theory after the condensation may be
constructed. Quevedo and Trugenberger considered as the only assumptions that the
resulting effective theory was renormalizable, Lorentz invariant, respects all the
symmetries of the problem and be constructed as a derivative expansion with respect
to the new scale defined by the characteristic density of the condensate.

In the present situation we are searching for a theory describing the physics of a
$P$ and $T$ symmetry breaking condensate. Taking as the starting point the scalar
field action (\ref{SNM}) we follow the Julia-Toulouse prescription by first identifying
the Stuckelberg-like brane symmetry (\ref{branesymm2}) invariant of the theory. This
is obviously given by:
\begin{eqnarray}
\label{braneinv}  f_{\mu}=\partial_{\mu} \phi - eK_{\mu}
\end{eqnarray}
The condensation process is a proliferation of the defects. A defect means a
singularity in the scalar field, that is, the scalar field is not well defined at
the position of a defect. As the condensation process becomes energetically favored
the scalar field becomes more and more singular until it is not defined anywhere and
only the brane invariant field $f_{\mu}$ (\ref{braneinv}) retains any physical
meaning. It describes the excitation field of the condensate. The Julia-Toulouse
prescription prompt us to add terms to the lagrangian to account for the dynamics of
these excitations. The first such term in a derivative expansion that breaks the $P$
and $T$ symmetries is the CS term. So we arrive at the following effective
description of the system after condensation of defects takes place
\begin{eqnarray}
\label{SD} {\cal L}_{SD} = \frac 12 f_{\mu}f^{\mu} +
\frac{1}{2m}f_{\mu}\varepsilon^{\mu\nu\rho}\partial_{\nu} f_{\rho}
\end{eqnarray}
where $m$ is interpreted as the density on the condensate.

We think that this is an important result. It gives not only a beautiful physical
picture of the otherwise \emph{ad hoc} deformation procedure in this instance, but
also constitutes a straightforward generalized application of the Julia-Toulouse
mechanism. Observe that this mechanism is here a dual representation for the
fermionic radiative processes, this suggests that there is more to the
Julia-Toulouse mechanism than originally thought.

\section{Conclusions and perspectives}

Besides providing a new representation of the well known $SD$ system through scalar
fields with deformed brackets, a nontrivial result by itself, the results described in this work provides a broad perspective on
the physical significance of the deformation procedure. It was
already suggested that the physical mechanism behind such deformations is associated
with radiative corrections \cite{Gamboa:2005bf}, and the deformation would be an emulation of
such processes. A well known deformation with this property is the Maxwell/MCS
deformation. The dual deformation obtained here, the scalar/SD deformation, must
thus be the emulation of a dual physical phenomenon. A remarkable property of this
new deformation is the phenomenon of ``rank jump'', that is, the change of the tensorial nature of the fundamental dynamical
fields in the theory. Using the fact that the phenomenon of ``rank jump'' may sometimes be associated with the condensation
of topological defects, a process better described by the so called Julia-Toulouse mechanism in the modern form discussed
by \cite{Quevedo:1996uu}, we were able to draw a physical picture of the deformation procedure in this $(2+1)D$ setting. An
interesting conclusion is the dual representation of the radiative processes. The known image of the vacuum as a medium with
quantum fluctuations of heavy particles modifying the propagation of lighter ones is reinterpreted as a condensate of defects
in which the dual of the lighter fields propagates.

\section{Acknowledgments}

Some of us, LG, MSG and CW would like to thank Funda\c{c}\~ao de Amparo \`a Pesquisa do Estado do
Rio de Janeiro (FAPERJ) and Conselho Nacional de Desenvolvimento Cient\' ifico e
Tecnol\'ogico (CNPq) and CAPES (Brazilian agencies) for financial support. JG and FM were  partially supported by FONDECYT-Chile grants 1050114, 1060079. CW and MSG would like to thank Departamento de F\'{\i}sica, Universidad de Santiago de Chile for kind hospitality and support.

\end{document}